\documentclass[aps,prc,twocolumn,superscriptaddress,nofootinbib,showpacs,showkeys,preprintnumbers]{revtex4-1}
\usepackage{graphicx}  
\usepackage{bm}        
\usepackage{amssymb}   
\usepackage{amsmath}   
\usepackage{hyperref}
\usepackage{multirow}
\usepackage{lineno}

\usepackage[usenames,dvipsnames]{xcolor}

\def\snn{\mbox{$\sqrt{s_{_{NN}}}$}}

\def\rap{\mathrm{y}}
\def\Nch{\mbox{$N_{\mathrm{ch}}$}}
\def\Npart{\mbox{$N_{\mathrm{part}}$}}
\def\Ncoll{\mbox{$N_{\mathrm{coll}}$}}
\def\dif{\mathrm{d}}

\begin{document}
\title{Next-generation multi-fluid hydrodynamic model for nuclear collisions at $\sqrt{s_{NN}}$ from few to hundered GeV}

\author{Jakub Cimerman} \affiliation{Faculty of Nuclear Sciences and Physical Engineering, Czech Technical University in Prague,\\  B\v rehov\'a 7, 11519 Prague 1, Czech Republic}
\affiliation{Univerzita Mateja Bela, Tajovsk\'eho 40, 974~01 Banská Bystrica, Slovakia}
\author{Iurii Karpenko} \affiliation{Faculty of Nuclear Sciences and Physical Engineering, Czech Technical University in Prague,\\  B\v rehov\'a 7, 11519 Prague 1, Czech Republic}
\author{Boris Tom\'a\v{s}ik} \affiliation{Faculty of Nuclear Sciences and Physical Engineering, Czech Technical University in Prague,\\  B\v rehov\'a 7, 11519 Prague 1, Czech Republic}
\affiliation{Univerzita Mateja Bela, Tajovsk\'eho 40, 974~01 Banská Bystrica, Slovakia}
\author{Pasi Huovinen}  \affiliation{Incubator of Scientific Excellence---Centre for Simulations of Superdense Fluids, University of Wroc\l{}aw, plac Maksa Borna 9, PL-50204 Wroclaw, Poland}

\begin{abstract}
We have developed a next-generation hybrid event-by-event three-fluid hydrodynamic model, suitable for simulations of heavy-ion collisions in the energy range from few up to tens of GeV per colliding $NN$ pair. At such energies the interpenetration time of the nuclei is of the same order as the lifetime of the system, however this model treats the initial phase hydrodynamically. Thanks to that it is more sensitive to the Equation of State than 1-fluid models with initial states being parametrised or generated by transport approach. Hence, our model is well designed for simulations at collision energies, at which matter in the vicinity of the QCD critical endpoint is expected. The construction of the model is explained and basic observables like hadron spectra in rapidity and transverse momentum, as well as elliptic flow are calculated.
\end{abstract}
\maketitle

\section{Introduction}

Ultra-relativistic heavy-ion collisions provide such conditions that nucleons melt into strongly interacting Quark-Gluon Plasma (QGP). Since its properties cannot be measured directly, we have to design phenomenological models to extract them by comparing the results of the model with the measured experimental data.

While LHC focuses on collisions at energies of few TeV, the energy range from few to few tens of GeV is no less interesting to study, especially because the critical endpoint is assumed to be probed in collisions at these energies. This energy range is currently being studied by the BES program at RHIC and the NA61/SHINE experiment at CERN, and two other facilities, NICA at JINR and FAIR at GSI, are under construction.

Hydrodynamic approach  has been one of the standard ways to simulate heavy-ion collisions since Landau and Bjorken \cite{Landau:1953gs, Bjorken:1982qr}. Today, pure hydrodynamic models have evolved into hybrid models, which combine hydrodynamic approach for the hot and dense stages of the collision evolution with a transport approach for final-state interactions.

Hydrodynamic modeling at RHIC BES energies is more challenging than at the top RHIC or LHC energies. The Lorentz contraction of colliding nuclei is much weaker. Therefore, the time of the interpenetration is of the same order as the lifetime of the hydrodynamic stage. This means that while some parts of the fireball are already in the hydrodynamic stage, in other parts nucleons are still approaching the collision zone. The picture of ``thin pancakes'' is no longer applicable, so one cannot assume a boost-invariant longitudinal expansion. It is also necessary to assume a finite baryon density of the produced medium. A hybrid model designed for top RHIC or LHC energies would not address these challenges and would not be suitable for energies lower than 20~GeV. 

The assumption of boost invariance has been relaxed and the finite baryon density included in several models in the literature. Studies of collisions at RHIC BES energies have been carried out using parametrized initial conditions
\cite{Cimerman:2020iny,Du:2022yok,Jiang:2023fad}, initial state from a transport model~\cite{Steinheimer:2014pfa,Karpenko:2015xea,Schafer:2021csj}, or so-called dynamical initialization~\cite{Shen:2022oyg,Akamatsu:2018olk} to take into account the complicated initial state geometry. However, when the initial state was modeled with a transport model, the hydrodynamic picture is applied after the complete passage of the incoming nuclei through each other, which means that a significant part of the evolution is modeled using hadronic degrees of freedom no matter how high the density. In the case of parametrized initial conditions, the starting time for the fluid stage was increased with the decrease of collision energy. When the initial time is large, there is reason to expect the transverse expansion to have started, but especially in event-by-event calculation it is difficult to provide physical constraints for parametrised transverse flow field. 

The dynamical initialization/fluidization approach avoids these caveats by treating each primary collision as a source term to the fluid, which can start evolving while the primary collisions are still going on. In this respect similar approach is the
so-called multi-fluid dynamics a.k.a.\ three-fluid dynamics. In this approach, the incoming nuclei are represented by two droplets of cold nuclear fluid, called projectile and target fluids. The process of heavy-ion collision is thus modeled as mutual interpenetration of the projectile and target fluids. The phenomenon of baryon stopping is modeled as friction between the projectile and target fluids. The kinetic energy lost to friction is channeled into creation of a third fluid, which represents particles produced in the reaction. Such concept of 3-fluid dynamics relies on fluid dynamical description of the heavy-ion collision from the very beginning. Assuming that even if the system is not close enough to local equilibrium for an EoS to be applicable, each one of these three subsystems---three fluids---is, this 
allows to model the compression stage of the reaction using fluid dynamics, and to probe its sensitivity to the EoS of dense nuclear matter.

Multi-fluid dynamical modeling of relativistic heavy-ion collisions has a long history, which can be rooted back to the two-fluid model of the Los Alamos group~\cite{Amsden:1978zz} and the later layout of a three-fluid formulation \cite{Csernai:1982zz}. In the 1990s and 2000s, a 3-fluid hydrodynamic model was developed by Mishustin, Russkikh and Satarov~\cite{Mishustin:1991sp} and subsequently improved by Ivanov, Russkikh and Toneev~\cite{Ivanov:2005yw}.

 The three-fluid dynamical model has been used to describe various observables, including transverse momentum spectra of various hadron species \cite{Ivanov:2018vpw}, directed flow \cite{Ivanov:2016sqy}, elliptic flow \cite{Ivanov:2014zqa}, light nuclei production \cite{Kozhevnikova:2020bdb}, or global $\Lambda$ polarization \cite{Ivanov:2022ble}. In the studies above, hadron distributions were evaluated via direct computation of Cooper-Frye integrals, and the hadronic phase was described using fluid dynamics, not hadron cascade.  Later, 3-fluid hydrodynamics was extended with UrQMD for final-state interactions, creating a hybrid model called THESEUS \cite{Batyuk:2016qmb}. However, this model has several shortcomings:
\begin{itemize}
    \item It lacks viscous corrections,
    \item hydrodynamic grid is defined in Cartesian coordinates, which is computationally inefficient in the presence of strong longitudinal expansion pertinent to collision energies $\snn>$20~GeV, 
    \item it lacks fluctuations of the initial state,
    \item EoS is hard-coded.
\end{itemize}

In this paper, we present MUFFIN\footnote{MUlti Fluid simulation for Fast IoN collisions}: a next-generation event-by-event three-fluid dynamic model based on the vHLLE code~\cite{Karpenko:2013wva}. MUFFIN is meant to be coupled to a final-state hadronic cascade. We use SMASH for this purpose \cite{Weil:2016zrk} forming MUFFIN-SMASH hybrid, which  addresses three out of the four above-mentioned issues and therefore is an ideal tool to simulate heavy-ion collisions. The present version of MUFFIN still relies on the perfect-fluid assumption, but we plan to include viscosity in near future. The technical description of the individual parts of MUFFIN is presented in Section \ref{sec:model}. Some general aspects of multi-fluid evolution are studied in Section \ref{sec:genprop} while the first results of our model are shown in Section \ref{sec:results}.


\section{The model}
\label{sec:model}
The multi-fluid evolution in hyperbolic coordinates 
\[
\tau = \sqrt{t^2-z^2}\, , \qquad \eta = \frac{1}{2} \ln \frac{t+z}{t-z}
\]
is solved using a modified vHLLE code \cite{Karpenko:2013wva}. An advantage of hyperbolic coordinates is that a fixed range in $\eta$ represents a volume which expands with the evolution time $\tau$, and one can simulate nucleus-nucleus collision at any $\snn$ with a hydrodynamic grid with fixed $\eta$ range. The pre-collision states of projectile and target fluids are constructed from randomly sampled coordinates of individual nucleons in the incoming nuclei. Thus, the initial states are fluctuating event-by-event. 
The hadrons are sampled at the hypersurface of particle-to-fluid transition, or particlization, using \texttt{SMASH-hadron-sampler} \cite{smash-hadron-sampler}, and final-state interactions simulated using the transport model SMASH \cite{Weil:2016zrk}. We describe the details of the model in the following.


\subsection{Initial state}
To account for event-by-event fluctuations, we construct the initial states of the projectile and target fluids by sampling the coordinates of individual nucleons inside the incoming nuclei, instead of assuming an average initial nuclear energy density. Local energy, momentum, baryon and electric charge densities of the fluids are then computed by smearing the point-like energies, momenta and charges of the nucleons in coordinate space using a smearing kernel. In this fashion, the evolution of the fireball is treated hydrodynamically from the very beginning.

We start by sampling the Cartesian coordinates of the nucleons inside the nuclei according to the Woods-Saxon formula \cite{Woods:1954zz}
\begin{equation}
\rho(x,y,z)=\frac{\rho_0}{1+\exp\left(\frac{\sqrt{x^2+y^2+z^2}-R}{a}\right)},
\label{eq:woods-saxon}
\end{equation}
where $a=0.459$~fm is a diffuseness  and
\begin{equation}
R=(1.1A^{1/3}-0.656A^{-1/3})~\mathrm{fm}    
\end{equation}
is the nuclear radius and $A$ is mass number of the nucleus. 
Next, since the fluid-dynamical evolution proceeds in hyperbolic coordinates, we have to set the incoming nuclei along the $\tau=\tau_0=\mathrm{const}$ hyperbola into a position before their first touch. 
\begin{itemize}
\item First, the generated positions of nucleons in $z$-coordinates are contracted (i.e.\ divided) by the $\gamma$ factor of the incoming nuclei $\gamma = \snn/2m_N = \cosh\rap{}$.
\item Then, the projectile (target) is moved to negative (positive) $z$ by $\zeta R/\gamma$. Here, $\zeta$ is a numerical factor chosen such that the nuclei do not overlap in the initial state. Its  values are 2 for energies $\snn = 7.7$~GeV and higher, but we also made runs at $\snn=3$~GeV and used $\zeta=1.1$ there. The time coordinate $t$ of each nucleon is set to $\tau_0$ at this point. Thus the collision technically starts at global time $t=\tau_0$ instead of $t=0$. We can do this because setting the clock is a matter of convention.
\item In the next step, all nucleons are free-propagated with the projectile (target) velocity $(-)\!\tanh\rap{}$ onto the $\tau = \tau_0$ hyperbola. Note that they do not encounter the other nucleus along this move, hence free propagation is adequate. 
\item The longitudinal positions of individual nucleons are distinguished by different values of $\eta$, as $z=\tau_0\sinh\eta$. The $\eta$ coordinates of the nucleons are calculated as
\end{itemize}
\begin{subequations}
\begin{align}
    \eta_{s,\mathrm{target}} &= \mathrm{asinh} \left(\frac{z}{\tau_0}\cosh\rap{}+\sinh\rap{}\right) - \rap{},\\
    \eta_{s,\mathrm{projectile}} &= \mathrm{asinh} \left(\frac{z}{\tau_0}\cosh\rap{}-\sinh\rap{}\right) + \rap{},
\end{align}
\end{subequations}
where $z$ is the original Cartesian coordinate and $\rap{}$ is the projectile rapidity. Finally, the nuclei are shifted along the $x$-direction by half of the impact parameter, to represent a non-central nucleus-nucleus collision.

Since the geometry of the system is neither based on two thin "pancakes" colliding, nor does the system depict the scaling flow $v_z = z/t$, we can freely choose the initial time $\tau_0$. As known, in hyperbolic coordinates even conservative algorithms tend to violate conservation laws~\cite{Karpenko:2013wva,Molnar:2014zha}. We have tested that varying $\tau_0$ from 0.75 to 5 fm/$c$ causes maximally $~3$\% change in total energy conservation with no visible change in other variables. The larger the $\tau_0$ the better the conservation laws are obeyed, and therefore we chose to use the value $\tau_0 = 5$ fm/$c$ which provides good conservation of energy (less than 1\% violation) and still allows most of the longitudinal expansion be captured by the expanding coordinate system.

Once the coordinates of the nucleons have been generated, they are transformed into fluids.  To smoothly deposit energies, momenta, baryon and electric charges of the incoming nucleons into hydrodynamic cells, we use a smoothing kernel from \cite{Oliinychenko:2015lva}:
\begin{align}
    &K(\Delta x, \Delta y, \Delta \eta_s) = \nonumber \\
    & A \exp\left(-\frac{\Delta x^2 + \Delta y^2 + \Delta \eta_s^2 \tau^2 \cosh^2\eta_s \cosh^2\rap{}}{2\sigma^2}\right), \nonumber
\end{align}
where $\Delta x, \Delta y, \Delta \eta_s$ represent the distance between a given nucleon and the center of a given fluid cell the energy is deposited into; $A$ is a numerically computed normalization constant so that the total energy, the baryon number, and the electric charge are conserved in the  procedure.
The energy-momentum densities as well as the 0th component of baryon and electric charge currents in each fluid cell are therefore summed up as follows:
$$T^{0\mu}(x_{\rm cell},y_{\rm cell},\eta_{\rm cell})=\sum\limits_{i \in \rm nucleons} p^\mu_i K(\Delta x, \Delta y, \Delta \eta_s)$$
$$N^{0}_b(x_{\rm cell},y_{\rm cell},\eta_{\rm cell})=\sum\limits_{i \in \rm nucleons} B_i K(\Delta x, \Delta y, \Delta \eta_s)$$
$$N^{0}_q(x_{\rm cell},y_{\rm cell},\eta_{\rm cell})=\sum\limits_{i \in \rm nucleons} Q_i K(\Delta x, \Delta y, \Delta \eta_s),$$
where $p^\mu$ is momentum of a hadron $i$, $B_i$ and $Q_i$ are its baryon and electric charges, respectively. Each hadron has the same longitudinal momentum with $p_i^0 = \snn/2$. At this stage we neglect the Fermi motion and set the transverse momentum of each hadron to zero. When decomposing the initial energy-momentum tensor and charge currents into densities and velocities, we assume no dissipative currents, and take pressure, temperature and chemical potentials according to the EoS.

An averaged initial state can also be constructed by generating a sample of initial states and taking averages of the energy-momentum tensor and the densities over the sample.

\subsection{Hydrodynamic evolution}

The projectile, target and fireball fluids coexist and partially overlap in the same coordinate space. The hydrodynamic evolution of individual fluids is computed in parallel using a modified vHLLE code~\cite{Karpenko:2013wva}. Although vHLLE has bulk and shear viscous corrections included, we keep them disabled in this work and leave viscous corrections in the multi-fluid picture for a future study.


\subsubsection{Interaction between fluids}
Local interaction between the fluids takes place as soon as more than one fluid is present in a given cell.
 Here, we follow the description by Ivanov et al.~\cite{Ivanov:2005yw}: The energy-momentum exchange between the fluids is given by  friction terms
\begin{subequations}
\begin{align}
\begin{split}
    \partial_\mu T^{\mu\nu}_\mathrm{p}(x)&=-F_\mathrm{p}^\nu(x)+F_\mathrm{fp}^\nu(x),
\end{split}\\
\begin{split}
    \partial_\mu T^{\mu\nu}_\mathrm{t}(x)&=-F_\mathrm{t}^\nu(x)+F_\mathrm{ft}^\nu(x),
\end{split}\\
\begin{split}
    \partial_\mu T^{\mu\nu}_\mathrm{f}(x)&=F_\mathrm{p}^\nu(x)+F_\mathrm{t}^\nu(x)-F_\mathrm{fp}^\nu(x)-F_\mathrm{ft}^\nu(x),
\end{split}
\end{align}
\end{subequations}
and there is no charge exchange between the fluids.
In the friction terms, the subscript denotes the fluid ($p$ stands for projectile, $t$ for target, and $f$ for fireball). The $F_{p}^\nu(x)$ and $F_{t}^\nu(x)$ are friction terms which correspond to projectile-target fluid friction and act upon the projectile and the target fluids, respectively. The $F_{fp}^\nu(x)$ and $F_{ft}^\nu(x)$ are friction terms which correspond to projectile-fireball and target-fireball friction. The friction terms for the fireball fluid are minus the sum of the friction terms for the projectile and target fluids, so that the total energy and momentum of the projectile, target and fireball fluids combined is conserved:
\begin{align}
    \partial_\mu \left[T^{\mu\nu}_{p}(x) + T^{\mu\nu}_{t}(x) + T^{\mu\nu}_{f}(x) \right]=0.
\end{align}
The friction between the projectile and the target fluids is parameterized as follows:
\begin{multline}
    F_\alpha^\nu=\vartheta^2 \rho_{p}^\xi \rho_{t}^\xi m_N V_{\mathrm{rel}}^{{pt}}[ (u_\alpha^\nu-u_{\overline{\alpha}}^\nu)\sigma_P(s_{{pt}})  \\  + (u_{p}^\nu+u_{t}^\nu)\sigma_E(s_{{pt}})]\, , 
    \label{eq:Falpha}
\end{multline}
where $m_N$ is the mass of the nucleon, $u_\alpha^\nu$ and $u_{\overline{\alpha}}^\nu$ are the 4-velocities of the fluids, with $\alpha$ index being $\alpha={p}$ or $t$, and the bar over the index means $\overline{{p}}={t}$ and $\overline{{t}}={p}$. 
The relative velocity of baryon-rich fluids $V_{\mathrm{rel}}^{{pt}}$ is  defined as
\begin{equation}
    V_{\mathrm{rel}}^{{pt}}=\frac{\sqrt{s_{pt}(s_{pt}-4m_N^2)}}{2m_N^2},
\end{equation}
where
\begin{equation}
s_{pt}=m_N^2(u_{p}^\nu+u_{t}^\nu)^2
\label{eq:spt}
\end{equation}
is the square of the mean invariant energy of the underlying colliding nucleons. Furthermore, $\vartheta$ is the overall factor depending on relative velocity, associated with the unification of the projectile and the target fluids when their relative velocity approach 0. It suppresses the friction exponentially
\begin{equation}
\vartheta = 1- \exp\left [ -(V_{\mathrm{rel}}^{{pt}}/\Delta V )^4\right ]\,  ,
\end{equation}
such that when the relative velocities of the fluids become small enough, the friction between them vanishes. Here, $\Delta V$ is the typical thermal velocity of particles within the fluid.

Other ingredients of Eq.~(\ref{eq:Falpha}) warrant a more thorough explanation:

$\blacksquare$ Scalars $\rho_\alpha^\xi$ represent effective densities of constituents of the projectile and target fluids in their respective rest frames. When the energy density of a fluid corresponds to hadronic phase, the fluid is dominated by baryons, therefore we equate $\rho_\alpha^\xi$ to net baryon density. When the energy density of a fluid corresponds to the quark-gluon phase, we associate the density with the sum densities of quarks, antiquarks and gluons. Furthermore, the sum is multiplied by a factor 1/3 to take into account that quarks and gluons have smaller cross-sections than nucleons, as predicted by the additive quark model \cite{Lipkin:1965fu}.  This leads to the following formula for  the effective density:
\begin{equation}
    \rho_\alpha^\xi(s_{pt})=
    \begin{cases}
    \rho_\alpha^b \xi_h(s_{pt}) &\; \varepsilon_\alpha < 0.7~\mathrm{GeV/fm}^3,\\
    \frac{1}{3}\left(\rho_\alpha^q+\rho_\alpha^g\right) \xi_q(s_{pt}) &\; \varepsilon_\alpha > 0.7~\mathrm{GeV/fm}^3.
    \end{cases}
\end{equation}
Here, $\rho_\alpha^b$, $\rho_\alpha^q$ and $\rho_\alpha^g$ are densities of net baryons, quarks, and gluons, respectively. Furthermore, we add scaling parameters (functions) $\xi_h$ and $\xi_q$, which will be discussed later in Sec.~\ref{sec:results-ft}. 
Note that the effective densities appear only in the friction terms, introduced below, and thus the $s_{pt}$-dependence can be actually attributed to the latter.
The quark and gluon densities are not evolved in the hydrodynamic code, therefore we reconstruct them using local temperature and baryon chemical potential in the limit of massless quarks and gluons \cite{Vogt:2007zz}:
\begin{subequations}
\begin{align}
\begin{split}
    \rho_\alpha^q &=\frac{18\zeta(3)}{\pi^2}T^3+2\mu_q^3,
\end{split}\\
\begin{split}
    \rho_\alpha^g &=\frac{16\zeta(3)}{\pi^2}T^3.
\end{split}
\end{align}
\end{subequations}
where the light-quark chemical potential is $\mu_q=\mu_{\rm B}/3$.

$\blacksquare$ $\sigma_{P/E}$ are cross-sections defined as
\begin{subequations}
\begin{align}
\begin{split}
    \sigma_P(s_{pt})&=\int_{\theta_{cm}<\pi/2}\mathrm{d}\sigma^{NN\rightarrow NX}\left(1-\cos \theta_{cm}\frac{p_{out}}{p_{in}} \right),
\end{split}\\
\begin{split}
    \sigma_E(s_{pt})&=\int_{\theta_{cm}<\pi/2}\mathrm{d}\sigma^{NN\rightarrow NX}\left(1-\frac{E_{out}}{E_{in}} \right).
\end{split}
\end{align}
\end{subequations}
In this way, $\sigma_P$ and $\sigma_E$ correspond to longitudinal momentum transport and energy transport, respectively.

The friction between baryon-rich and fireball fluid is given by
\begin{equation}
    F_{{f}\alpha}^\nu = \rho_\alpha^b\xi_{{f}\alpha}(s_{{f}\alpha}) V_{\mathrm{rel}}^{{f}\alpha}\frac{T_{{f}(eq)}^{0\nu}}{u_{f}^0}\sigma_{tot}^{N\pi\rightarrow R}(s_{{f}\alpha}), \label{eq:Ffalpha}
\end{equation}
where $\xi_{{f}\alpha}(s_{{f}\alpha})$ is the tuning parameter, 
\begin{equation}
s_{{f}\alpha}=(m_\pi u_{f}+m_Nu_\alpha)^2
\label{eq:sfalpha}
\end{equation}
and $V_{{rel}}^{{f}\alpha}$ is the mean invariant relative velocity between baryon-rich and fireball fluids defined as
\begin{equation}
    V_{\mathrm{rel}}^{{f}\alpha}=\frac{\sqrt{(s_{{f}\alpha}-m_N^2-m_\pi^2)^2-4m_N^2m_\pi^2}}{2m_Nm_\pi}.
\end{equation}

\subsection{Equation of state}

An advantage of MUFFIN, inherited from the basic vHLLE code, lies in the possibility of changing the equation of state. Thanks to that, the model can be used to study the sensitivity of various observables to the EoS. However in this paper, for the general benchmark of the model we use only one EoS based on an effective chiral hadron-quark model \cite{Steinheimer:2010ib} that qualitatively matches to lattice QCD results at $\mu_B=0$ and to hadron-resonance gas with excluded volume corrections at low temperatures. This EoS has an advantage of being defined in the whole $T$--$\mu_B$ plane, and is used for the evolution of all fluids; however, its low-temperature limit quantitatively differs from the effective hadron-resonance gas EoS in SMASH; therefore, following a recipe from \cite{Cheng:2010mm}, for the computation of flow velocity, temperature and chemical potentials at the particlization hypersurface from energy-momentum density at it, we use hadron-resonance gas EoS from SMASH \cite{Schafer:2021rfk}. This ensures that the energy, momentum and quantum numbers are conserved in the particlization process.


\subsection{Fluid-to-particle transition and final-state interactions}

In hybrid models for top RHIC or LHC energies, one typically assumes that the fluid-to-particle transition, or particlization, takes place at a fixed temperature, which should be low enough so that the medium is locally in hadronic phase, and in the range where the fluid-dynamical and transport descriptions are both valid. At lower collision energies, where the effects of baryon density become non-negligible, 
the phase transition temperature decreases, and the use of the same particlization temperature as at high energies is not advisable. To avoid adjusting the particlization temperature for each collision energy, it is practical to use constant energy density as particlization criterion.

With more than one fluid in the picture, a proper particlization criterion is more ambiguous. If each fluid particlizes individually, space-time regions will appear with a mixture of fluid and particles, which complicates the modeling. To avoid such complications, we choose to particlize all fluids at the same hypersurface in space-time. For the particlization criterion, we choose  a fixed ``combined'' energy density of $\varepsilon_{{sw}} = 0.5$ GeV/fm$^3$. To compute the latter, at each space-time cell, we take the combined energy-momentum tensor of all fluids, $T^{\mu\nu}_{p}(x) + T^{\mu\nu}_{t}(x) + T^{\mu\nu}_{f}(x)$, and diagonalize it to extract such combined energy density as its first eigenvalue. With a field of combined energy-density in space-time, Cornelius subroutine \cite{Huovinen:2012is} is used to construct the particlization hypersurface. The constructed hypersurface is composed of many small segments.

\begin{figure}[h]
    \centering
    \includegraphics[width=0.49\textwidth]{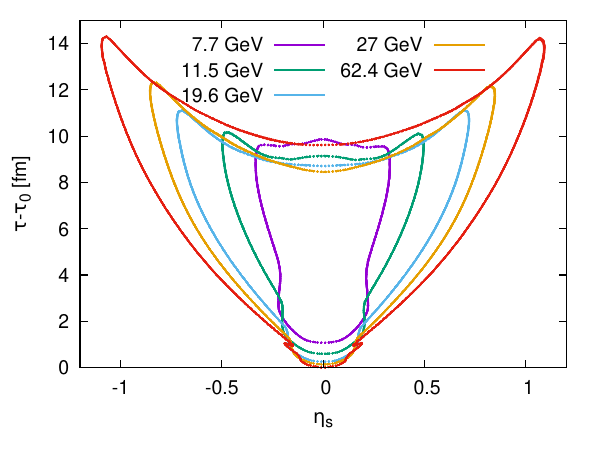}
    \caption{$\eta_s-\tau$ sections of particlization hypersurface at $r_T=0$, constructed in simulations of central Au-Au collisions for different collision energies with averaged initial state.} \label{fig:particlization}
\end{figure}

\begin{figure}[h]
    \centering
    \includegraphics[width=0.49\textwidth]{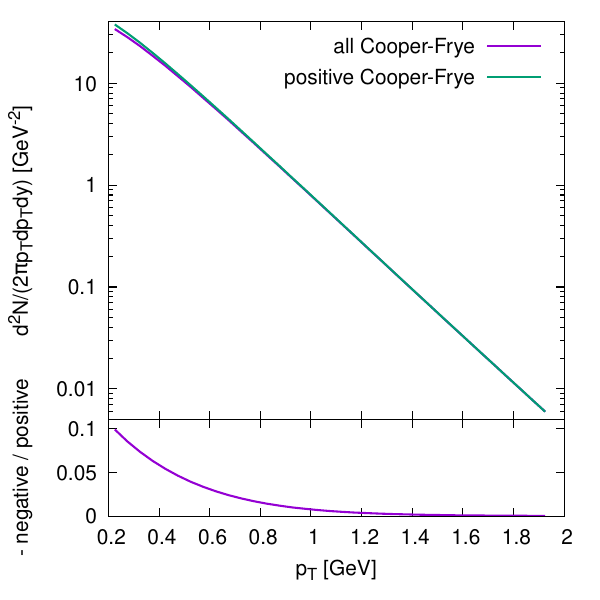}
    \caption{Positive and positive+negative Cooper-Frye contributions to the  thermal pion yield at midrapidity as a function of pion transverse momentum (top panel) and the ratio of minus negative to positive contributions (bottom panel) as a function of pion transverse momentum, computed in a multi-fluid simulation of central Au-Au collision at $\snn=7.7$~GeV.}
    \label{fig:negative_cooper_frye}
\end{figure}

In conventional one-fluid calculations, most of the system at the initial time of fluid-dynamic evolution, $\tau_0$, is hot and within the particlization hypersurface. The initial state of multi-fluid calculation consists of cold nuclear matter, and therefore no part of the system is initially within the particlization hypersurface. Consequently the constant energy density hypersurface forms an enclosed surface as demonstrated in Figure \ref{fig:particlization}.

As per Gauss' theorem, the net energy and momentum flows through enclosed surface must be zero, 
$$\int d\Sigma_\mu T^{0\mu}=0,$$
and consequently there are regions on the hypersurface where energy and momentum flows through the surface are negative, i.e.~directed inwards. These are the regions where the initial state matter is flowing inwards, towards the hot and dense interaction region, and must therefore be excluded from the calculation of final state particles at particlization hypersurface. We filter out such hypersurface segments based on the following criteria:
\begin{subequations}
\begin{align}
    \mathrm{d}\Sigma^\mu \mathrm{d}\Sigma_\mu > 0 \; &\mathrm{and}\; \mathrm{d}\Sigma_0 <0, \\
    \mathrm{d}\Sigma^\mu \mathrm{d}\Sigma_\mu < 0 \; &\mathrm{and}\; \mathrm{d}\Sigma_\mu T^{\mu 0} 
    < 0,
\end{align}
\end{subequations}
where $\mathrm{d}\Sigma^\mu$ is the normal vector of the hypersurface and 
$\mathrm{d}\Sigma_\mu T^{\mu 0}$ is the energy flow through the hypersurface. In numerical calculations the requirement that the net energy flow through enclosed surface is zero can be used to check the accuracy of the calculations. We have checked that in our calculations the net flow of energy is less than 5\% of the total outflow of energy through the constant density hypersurface.

It is known that if the hypersurface is spacelike, the Cooper-Frye procedure allows negative contributions to the particle distributions. Even after removing the segments of the hypersurface where the energy flow is directed inwards, significant part of the surface is spacelike, cf.~Figure~\ref{fig:particlization}. To check whether negative Cooper-Frye contributions might be a problem in our model, we show in Fig.~\ref{fig:negative_cooper_frye} positive, positive+negative contributions and the ratio of negative to positive contributions to the $p_T$ spectrum of thermal pions  at the particlization surface in a multi-fluid simulation of a central Au-Au collision at $\snn=7.7$~GeV with an averaged initial state. The contributions were computed by a direct numerical integration of the Cooper-Frye formula (see below). One can see that the negative Cooper-Frye contribution  is relatively small at very low $p_T$, and becomes negligible as the $p_T$ increases.

In hybrid MUFFIN-SMASH calculations, hadrons are sampled at the  particlization hypersurface according to the Cooper-Frye formula \cite{Cooper:1974mv}
\begin{equation}
    N=\int\frac{\mathrm{d}^3p}{E_p}\int \mathrm{d}\Sigma_\mu(x)p^\mu f(p, T(x), \mu_i(x)).
\end{equation}
The sampling process is carried out using the \texttt{SMASH-hadron-sampler} \cite{smash-hadron-sampler}, with the details of sampling algorithm described in \cite{Karpenko:2015xea}. The sampling algorithm includes viscous corrections to hadron distribution functions, however for the present study they are not relevant as viscosity is switched off in the hydro stage. The final-state interactions are then simulated with the microscopic transport model SMASH \cite{Weil:2016zrk}, which includes resonance decays, 2-particle inelastic and elastic scatterings, and resonance excitations.


\section{General properties of multi-fluid evolution}
\label{sec:genprop}

\begin{figure*}
 \centering
 \includegraphics[width=0.32\textwidth]{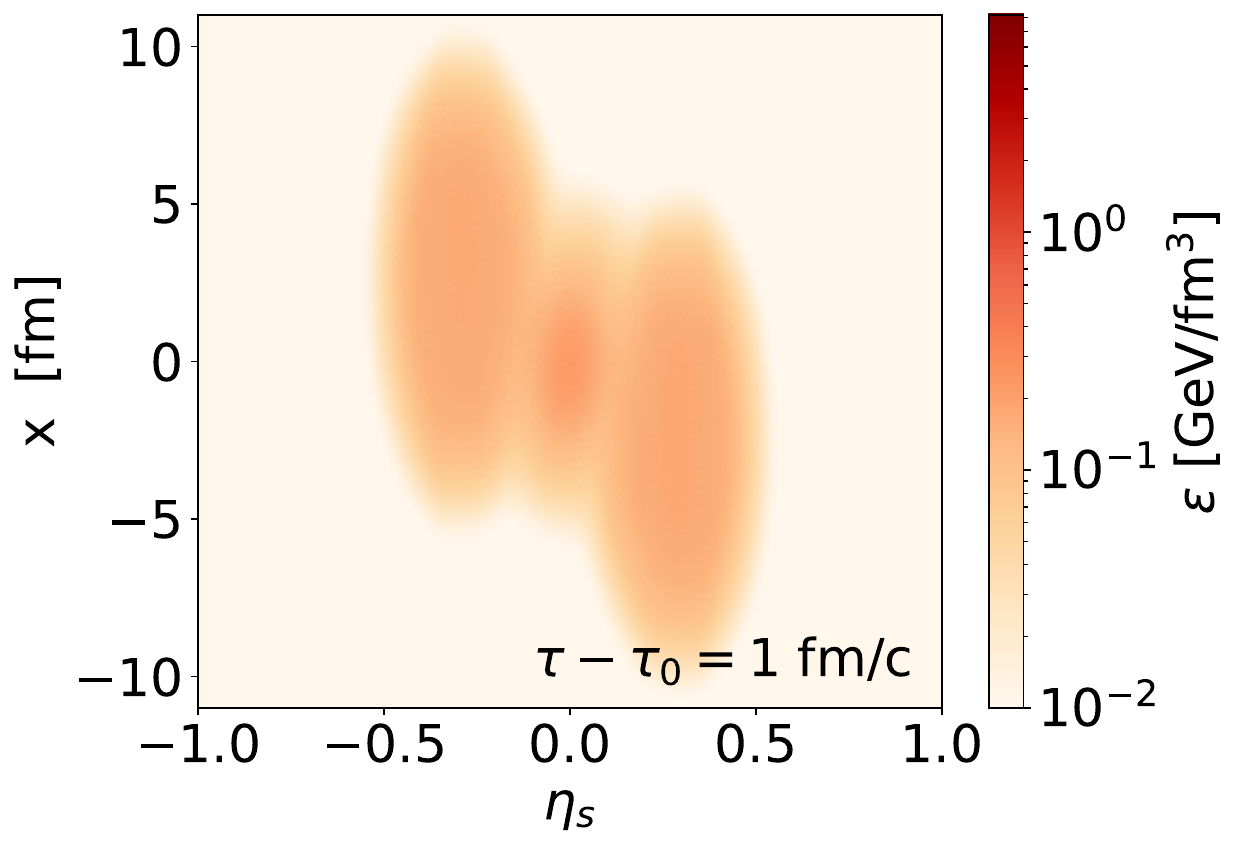}
 \includegraphics[width=0.32\textwidth]{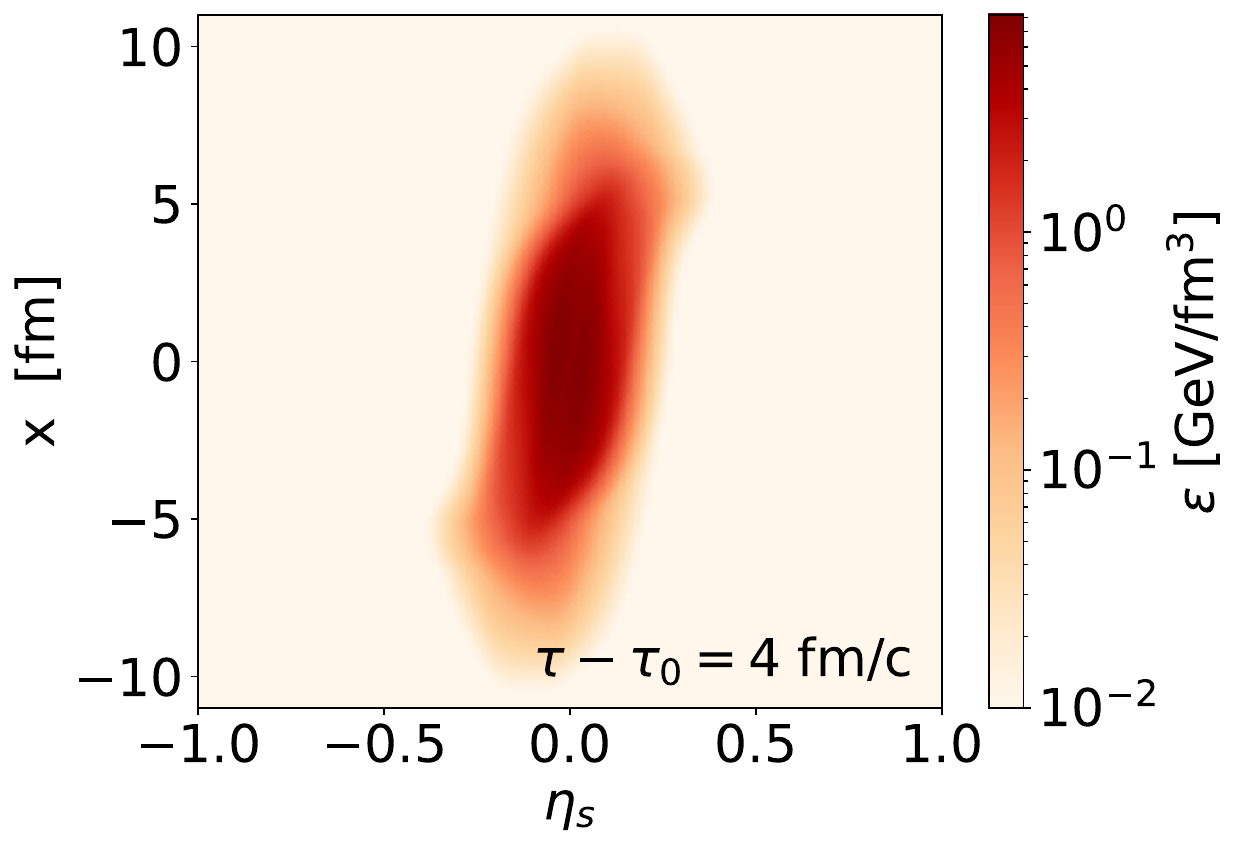}
 \includegraphics[width=0.32\textwidth]{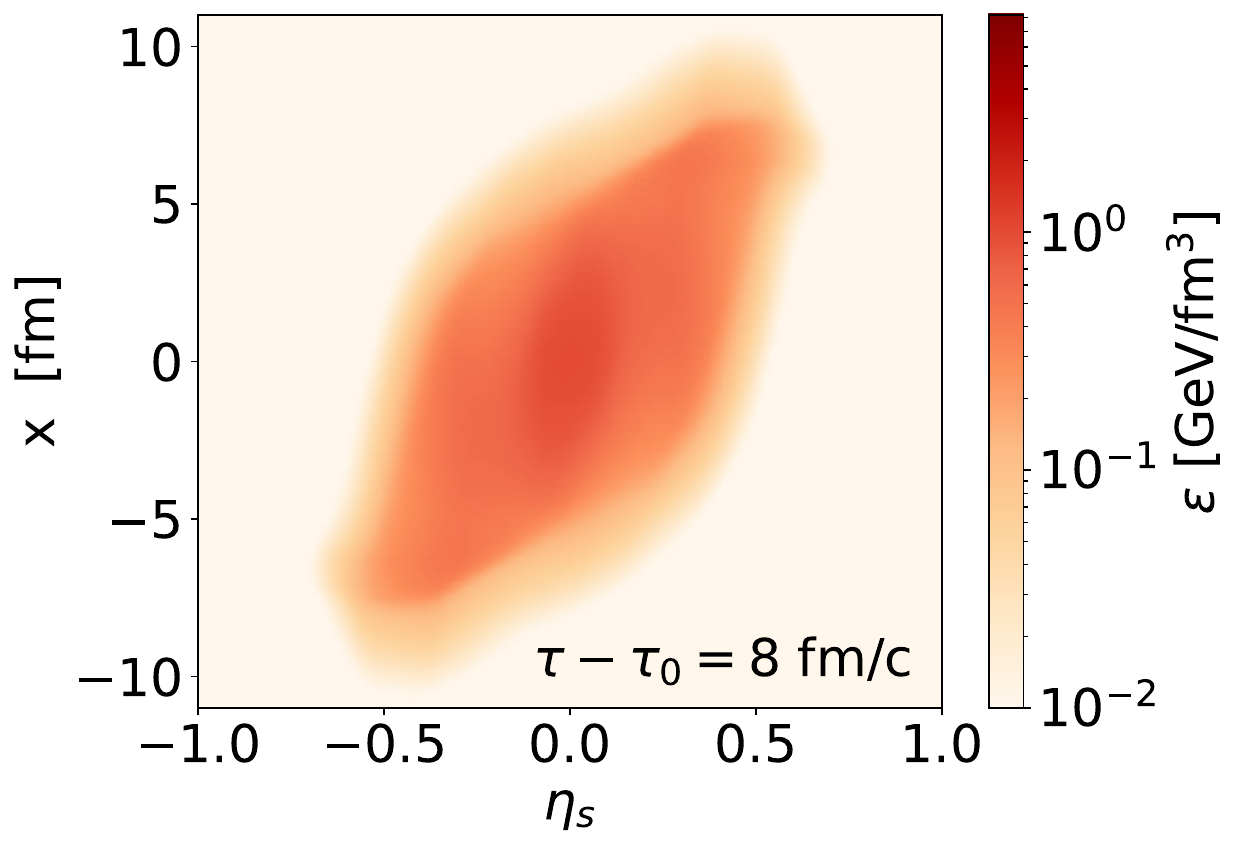}
 \caption{Distributions of combined energy density of the fluids in $x-\eta_s$ plane at $y=0$. Three different stages of evolution of a Au+Au collision at $\snn = 7.7$~GeV are displayed. The text labels show $\tau-\tau_0$, time after the beginning of interpenetration of the fluids.}\label{fig:fluids_evolution}
\end{figure*}

\begin{figure}
 \centering
 \includegraphics[width=0.5\textwidth]{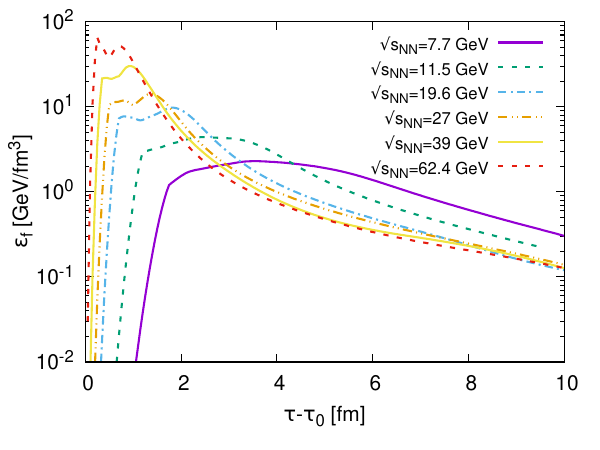}
 \caption{Time evolution of the energy density in the central cell ($x=y=\eta_s=0$) of the fireball fluid, in the simulations of central Au-Au collisions at different collision energies in the BES range.}\label{fig:central_cell}
\end{figure}

\begin{figure}
 \centering
 \includegraphics[width=0.5\textwidth]{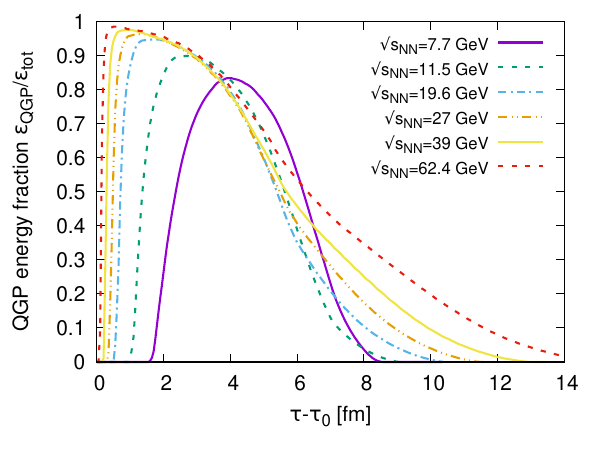}
 \caption{Time evolution of the fraction of medium in the QGP phase, in the simulations of central Au-Au collisions at different collision energies in the BES range.}\label{fig:qgp_fraction}
\end{figure}

We start by examining the basic properties of the multi-fluid evolution at different BES energies. For that purpose, the simulations were conducted with averaged initial states, i.e.\ when the initial energy and charge distributions in the projectile and target fluids were averaged from many sampled distributions of the projectile and target nucleons.

Figure~\ref{fig:fluids_evolution} shows distributions of the  combined energy density of the fluids, which is obtained after the diagonalization of the combined energy-momentum tensor $T^{\mu\nu}_{p}(x) + T^{\mu\nu}_{t}(x) + T^{\mu\nu}_{f}(x)$. The left panel represents an early stage of collision, 1 fm/c after the initialisation of the nuclei, where one can see the fireball fluid starting to form in the middle. The central panel shows the most dense stage of evolution, with the combined energy density raising up to several GeV/fm$^3$. The right panel shows the late stage of evolution, with parts of projectile and target fluids flying away, and fireball fluid expanding and cooling down.

Figure~\ref{fig:central_cell} shows time evolution of the energy density in the central ($x=y=\eta_s=0$) cell of the fireball fluid. The first observation from this plot is that the cell starts to heat up later as the collision energy decreases. Even at the lowest collision energy $\snn=7.7$~GeV the incoming fluids move with relativistic velocities. Nevertheless, due to the weaker Lorentz contraction of the fluids, it takes longer for the fluids to reach the state of maximal overlap, when the friction is strongest. At the highest collision energy, a double-peaked structure starts to develop in the time evolution of the energy density. The latter happens due to fireball-projectile and fireball-target friction, which starts to act later than the projectile-target friction, and draws energy from the hotter fireball fluid  back to less hot projectile/target fluids.

The maximal energy density of the central cell decreases dramatically with decreasing $\snn$, however, the lifetime of the dense (QGP) phase of matter, which we define as $\varepsilon>0.5$~GeV/fm$^3$, somewhat increases in the central cell. The longer lifetime of the dense phase is a consequence of a longer interpenetration phase of the projectile and target fluids, and less violent expansion dynamics. We define the fraction of the QGP phase at a given time as a fraction of the total energy of the system carried by fluid cells with local energy density $\varepsilon>0.5$~GeV/fm$^3$:
\begin{equation}
  \frac{\varepsilon_{QGP}}{\varepsilon_{tot}}=\frac{\sum_{i=\rm p,t,f} \int \dif\eta\,\dif^2 r_\perp\, T^{00}_i\,\theta(\varepsilon_i-\varepsilon_{{sw}})}{\sum_{i=\rm p,t,f} \int \dif\eta\,\dif^2 r_\perp\,T^{00}_i},
\end{equation}
where $\varepsilon_{{sw}} = 0.5$ GeV/fm$^3$. 
As seen in Figure~\ref{fig:qgp_fraction}, the maximum value of the QGP fraction 
slightly decreases with decreasing $\snn$ but stays quite high even for the lowest collision energy simulated, $\snn=7.7$~GeV. This is also confirmed in the middle panel of Figure~\ref{fig:fluids_evolution}. The large QGP fraction at all considered collision energies is a result of the friction, which relatively easily converts the kinetic energy of the projectile and target into the internal energy of the fireball fluid.

Note that the evolution of the QGP energy fraction $\varepsilon_{\mathrm{QGP}}/\varepsilon_{\mathrm{tot}}$ has also been calculated within the PHSD model \cite{Moreau:2021clr}. While in our simulations the system always passes through a state where the ratio is at least 0.8, PHSD predicts that the maximum value drops from 0.9 at $\snn=200$~GeV down to about 0.25 at $\snn=7.7$~GeV. We attribute the discrepancy with QGP fraction in PHSD in part to a different method to count the fraction, which is computed in PHSD as a ratio of the mean energy of partons over the sum of mean energies of partons, baryons and mesons at mid-rapidity \cite{PHSD-QGP-fraction}.


\section{Results}
\label{sec:results}

In this section, we present the first results from MUFFIN-SMASH, the developed three-fluid hybrid  model with event-by-event fluctuating initial conditions. We simulated Au+Au collisions at 6 RHIC BES energies: $\snn = 7.7$, $11.5$, $19.6$, $27$, $39$, and $62.4$~GeV. For each energy, we have run 3000 hydrodynamic simulations. To increase the statistics, we oversampled hadrons and produced 500 final-state events from each of the 3000 hydrodynamic configurations.

\subsection{Fine-tuning}
\label{sec:results-ft}

The friction terms represent the biggest unknown in the model. As there is no rigorous derivation of the friction terms from the underlying kinetic theory, equations \eqref{eq:Falpha} and \eqref{eq:Ffalpha} can be considered as reasonable assumptions about the functional form, and the  dependence of the friction on the relative velocity. As such, there is certain freedom with both the shape and the strength of the friction terms, and we treat those terms essentially as fitting parameters, fixing them from model-to-data comparison.

The strength of the friction terms is regulated using the scaling parameters $\xi_h$, $\xi_q$, and $\xi_{{f}\alpha}$. The parameters control the strength of baryon stopping and the amount of energy-momentum transferred from the baryon-rich fluids to the fireball fluid. The model was tuned on transverse momentum spectra and rapidity distributions of net-protons at available collision energies. We found that to optimally reproduce the observables at different collision energies, the scaling parameters had to change with $\snn$. However, as the fluid cells do not know about the global colliding energy, we chose the friction scaling to depend on the invariant energy of the colliding fluid elements. We tried several polynomial dependencies on the invariant energy and found that the experimental data for all studied energies were reproduced best with the following parameter values, which were used for the calculations presented in this paper:
\begin{subequations}
\begin{align}
    \xi_h = 1.8\sqrt{\frac{2m_N}{\sqrt{s_{pt}}}},\\
    \xi_q = 30\sqrt{\frac{2m_N}{\sqrt{s_{pt}}}},\\
    \xi_{{f}\alpha} = 0.15\frac{m_N^2}{s_{{f}\alpha}},
\end{align}
\end{subequations}
where $s_{pt}$ was defined in eq.~(\ref{eq:spt}) and $s_{{f}\alpha}$ was defined in eq.~(\ref{eq:sfalpha}).

\subsection{Centrality determination}

In this paper, we mostly use data from the STAR experiment. They use pseudorapidity density of charged hadrons $dN_{\rm ch}/d\eta$ at mid-rapidity as a measure of centrality\footnote{Technically, a so-called raw multiplicity $dN_{\rm ch}^{\rm raw}/d\eta$ is used, which does not include corrections for trigger, acceptance and detector inefficiencies}, see e.g.\ \cite{STAR:2013ayu}. Following the same definition in our studies is not straightforward - most importantly, to our knowledge, STAR does not publish exact ranges in $dN_{\rm ch}/d\eta$ for the different centrality classes. Therefore, to follow the STAR definition we would need to simulate minimum-bias events in MUFFIN-SMASH, make sure that the multiplicity distribution is compatible with the experiment, then bin the generated events into different centrality classes using $dN_{\rm ch}/d\eta$. However, application of fluid-dynamical approach to peripheral heavy-ion collisions is challenging, and we do not expect the multi-fluid model to reproduce the experimental data well in that regime. Nevertheless, we prefer to avoid using proxy measures such as ranges in impact parameter or number of participants, and conducted the following procedure for centrality selection.

We generated events with impact parameters in the range 0-12~fm, which approximately corresponds to 0-50\% centrality. Then, we simulated the multiplicity distribution in minimum-bias scenario using a two-component model \cite{Kharzeev:2000ph}. In this model, the multiplicity in nuclear collisions has contributions from the ``soft'' part, which is proportional to the mean number of participants $\langle\Npart\rangle$, and from the ``hard'' part, which is proportional to the mean number of binary collisions $\langle\Ncoll\rangle$
\begin{equation}
    \frac{\mathrm{d}\Nch}{\mathrm{d}\eta}=n_{pp}\left[ (1-x)\frac{\langle\Npart\rangle}{2}+x\langle\Ncoll\rangle\right].
\end{equation}
Here, $n_{pp}$ is the average multiplicity in minimum-bias p+p collisions, and $x$ is the fraction of the hard component. In our procedure, we first simulated event-by-event \Npart{} and \Ncoll{} using the Monte Carlo Glauber (MCG) model \cite{Loizides:2017ack}. With those numbers we determined 
\[
M = \left[ (1-x)\frac{\Npart}{2}+x\Ncoll\right]
\]
which was then rounded to become integer. Here, $x=0.11$ was chosen \cite{STAR:2009sxc}. In the next step, we convoluted $M$ times the negative binomial distribution (NBD) 
\begin{equation}
    P_{\text{NBD}}(n_{pp},k;n)=\frac{\Gamma(n+k)}{\Gamma(n+1)\Gamma(k)}\frac{(n_{pp}/k)^n}{(n_{pp}/k+1)^{n+k}}
\end{equation}
to produce the final multiplicity as a sum of $n$'s  from the individual NBDs. The value $k=2.1$ was used, following \cite{STAR:2009sxc}. The value of $n_{pp}$ was obtained by fitting the multiplicity distribution from the 3-fluid model (see Table \ref{tab:glauber_parameters}). 


\begin{table}[tb]
 \centering
 \begin{tabular}{ c | c | c }
 $\snn$ [GeV] & $\sigma_{NN}$ [mb] & $n_{pp}$ \\ \hline \hline
 $7.7$ & $30.6$ & $0.89$ \\ 
 $11.5$ & $31.28$ & $0.99$ \\ 
 $19.6$ & $32.3$ & $1.11$ \\
 $27$ & $33.1$ & $1.16$ \\
 $39$ & $34.2$ & $1.23$ \\
 $62.4$ & $35.9$ & $1.36$ \\
 \end{tabular}
\caption{Parameters of the two-component MCG model - inelastic nucleon-nucleon cross-section $\sigma_{NN}$, and the average multiplicity in minimum-biased p+p collisions $n_{pp}$, for BES energies.}
\label{tab:glauber_parameters}
\end{table}

\begin{figure}
    \centering
    \includegraphics[width=0.5\textwidth]{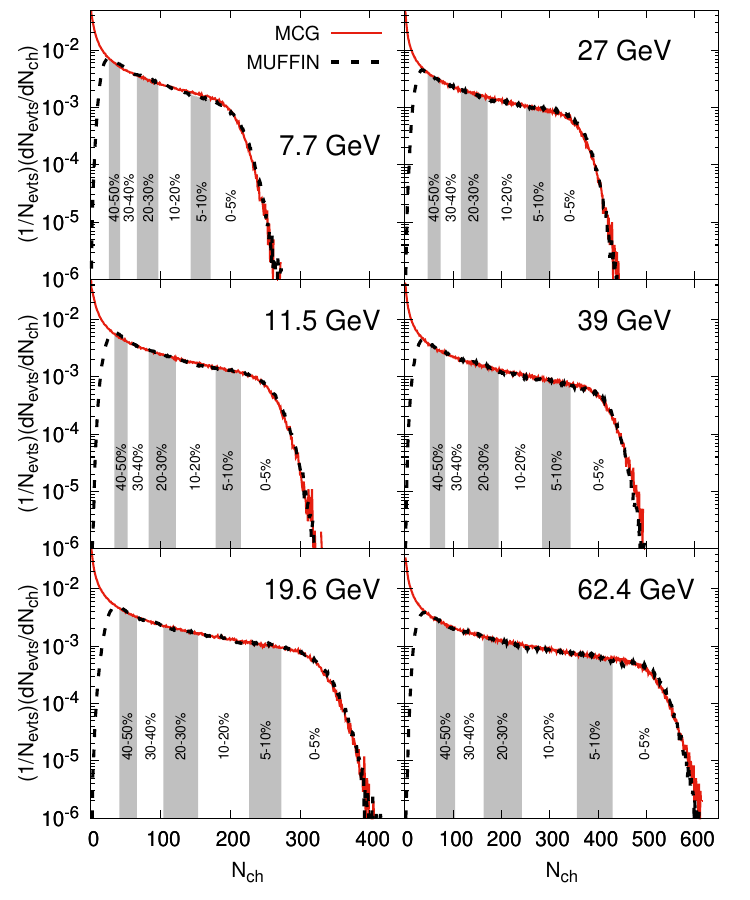}
    \caption{Multiplicity distributions for BES energies $\snn=7.7$, $11.5$, $19.6$, $27$, $39$, and $62.4$~GeV obtained from the hybrid three-fluid model MUFFIN-SMASH (dashed black curves) compared with MCG model (solid red curves). This comparison is used to determine centrality classes in our model, and they are illustrated with gray and white areas.}
    \label{fig:centrality_determination}
\end{figure}

Finally, we scaled the multiplicity distribution from our model with the ratio of the number of events with $\Nch{}>50$ obtained from the MCG simulation to the same obtained from our model. This results in very well-reproduced multiplicity distributions (see Fig. \ref{fig:centrality_determination}), which we used to obtain the multiplicity ranges for the determination of the centrality. These ranges are listed in Table \ref{tab:centrality_determination}, and for $\snn=62.4$~GeV they are consistent with the mean multiplicities for the different centrality classes published by STAR \cite{STAR:2008med}.

\begin{table}[tbh!]
    \centering
    \begin{tabular}{cc|ccc}
    Centrality & $\mathrm{d}\Nch/\mathrm{d}\eta$ & $b$ [fm] & $\langle \Npart\rangle$ & $\langle\Ncoll\rangle$ \\ \hline\hline
    \multicolumn{5}{c}{Au+Au $7.7$~GeV} \\ \hline
    $0-5\%$ & $\geq 172$ & $0.00-3.17$ & $336.7$ & $774.7$ \\
    $5-10\%$ & $142-171$ & $3.17-4.54$ & $287.8$ & $626.6$ \\
    $10-20\%$ & $97-141$ & $4.54-6.46$ & $223.9$ & $450.1$ \\
    $20-30\%$ & $65-96$ & $6.46-7.93$ & $157.9$ & $283.1$ \\
    $30-40\%$ & $42-64$ & $7.93-9.16$ & $108.6$ & $171.5$ \\
    $40-50\%$ & $25-41$ & $9.16-10.31$ & $70.6$ & $96.5$ \\ \hline
    \multicolumn{5}{c}{Au+Au $11.5$~GeV} \\ \hline
    $0-5\%$ & $\geq 214$ & $0.00-3.20$ & $338.2$ & $793.6$ \\
    $5-10\%$ & $177-213$ & $3.20-4.56$ & $288.3$ & $638.6$ \\
    $10-20\%$ & $121-176$ & $4.56-6.47$ & $224.6$ & $458.9$ \\
    $20-30\%$ & $81-120$ & $6.47-7.94$ & $158.4$ & $288.0$ \\
    $30-40\%$ & $52-80$ & $7.94-9.19$ & $108.7$ & $173.6$ \\
    $40-50\%$ & $32-51$ & $9.19-10.26$ & $71.5$ & $99.0$ \\ \hline
    \multicolumn{5}{c}{Au+Au $19.6$~GeV} \\ \hline 
    $0-5\%$ & $\geq 273$ & $0.00-3.20$ & $340.3$ & $821.7$ \\
    $5-10\%$ & $226-272$ & $3.20-4.57$ & $290.1$ & $660.1$ \\
    $10-20\%$ & $154-225$ & $4.57-6.48$ & $225.9$ & $473.2$ \\
    $20-30\%$ & $103-153$ & $6.48-7.95$ & $159.2$ & $296.0$ \\
    $30-40\%$ & $66-102$ & $7.95-9.20$ & $109.1$ & $177.6$ \\
    $40-50\%$ & $40-65$ & $9.20-10.32$ & $71.3$ & $100.0$ \\ \hline
    \multicolumn{5}{c}{Au+Au $27$~GeV} \\ \hline
    $0-5\%$ & $\geq 301$ & $0.00-3.21$ & $341.3$ & $842.2$ \\
    $5-10\%$ & $249-300$ & $3.21-4.59$ & $290.7$ & $674.3$ \\
    $10-20\%$ & $170-248$ & $4.59-6.49$ & $226.7$ & $483.9$ \\
    $20-30\%$ & $114-169$ & $6.49-7.95$ & $160.5$ & $303.6$ \\
    $30-40\%$ & $73-113$ & $7.95-9.20$ & $110.0$ & $182.0$ \\
    $40-50\%$ & $45-72$ & $9.20-10.28$ & $72.4$ & $103.2$ \\ \hline
    \multicolumn{5}{c}{Au+Au $39$~GeV} \\ \hline
    $0-5\%$ & $\geq 343$ & $0.00-3.22$ & $342.8$ & $870.6$ \\
    $5-10\%$ & $283-342$ & $3.22-4.62$ & $291.4$ & $694.5$ \\
    $10-20\%$ & $194-282$ & $4.62-6.49$ & $227.7$ & $498.0$ \\
    $20-30\%$ & $130-193$ & $6.49-7.96$ & $161.3$ & $312.4$ \\
    $30-40\%$ & $83-129$ & $7.96-9.22$ & $110.7$ & $187.1$ \\
    $40-50\%$ & $51-82$ & $9.22-10.30$ & $72.9$ & $105.7$ \\ \hline
    \multicolumn{5}{c}{Au+Au $62.4$~GeV} \\ \hline
    $0-5\%$ & $\geq 429$ & $0.00-3.23$ & $344.9$ & $915.1$ \\
    $5-10\%$ & $354-428$ & $3.23-4.63$ & $293.7$ & $728.5$ \\
    $10-20\%$ & $241-353$ & $4.63-6.53$ & $229.1$ & $520.4$ \\
    $20-30\%$ & $161-240$ & $6.53-8.00$ & $162.0$ & $323.9$ \\
    $30-40\%$ & $103-160$ & $8.00-9.25$ & $111.4$ & $193.8$ \\
    $40-50\%$ & $63-102$ & $9.25-10.34$ & $73.3$ & $108.6$ \\ 
    \end{tabular}
    \caption{Limits of multiplicities $\Nch$ within $|\eta|<0.5$ used for centrality determination, and impact parameter range, $\langle \Npart\rangle$, and $\langle \Ncoll\rangle$ extracted from the MCG model for BES energies and centralities $0-50\%$.}
    \label{tab:centrality_determination}
\end{table}

\begin{figure}
    \centering
    \includegraphics[width=0.5\textwidth]{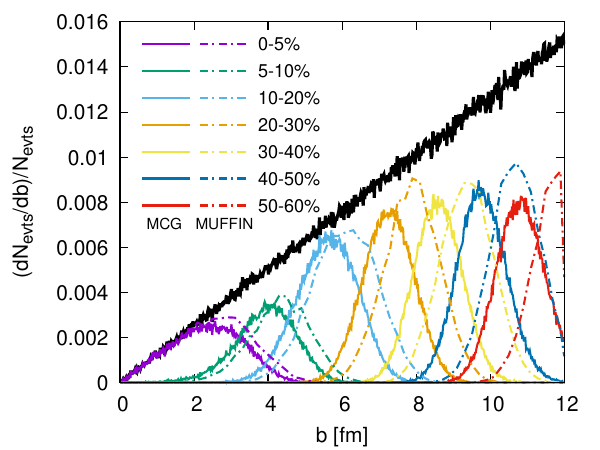}
    \caption{Histogram of impact parameters for various centralities in Au+Au collisions at $\snn=7.7$~GeV obtained from MCG model (solid lines) and hybrid three-fluid model MUFFIN-SMASH (dash-dotted lines). The solid black line is the histogram of the impact parameter in minimum-biased MCG simulations.}
    \label{fig:b_histogram}
\end{figure}

The impact parameter, \Npart{}, and \Ncoll{} in Table \ref{tab:centrality_determination} are just informative and do not play any role in determining the centrality bins in our model. However, an interesting finding is that the impact parameter ranges obtained from our model differ from those obtained from the MCG model. This is illustrated in Fig. \ref{fig:b_histogram}. The MCG model assumes smaller values of the impact parameters for the same \Nch{} as compared to our model. This means that if we had used the impact parameter ranges from the MCG model to define centrality classes, we would have overestimated the multiplicities of hadrons. This discrepancy is rooted in the Glauber model, which is a purely geometrical model where the nucleons propagate along straight lines even after interacting, and there is a sharp separation between participant and spectator nucleons. However, when the Lorentz contraction of the projectile and the target is weak, and the interpenetration takes a relatively long time, the produced fireball, as well as participant parts of the projectile and target, start to expand when the spectators are still around the interaction region. Therefore, the spectators, as defined by the Glauber model, can and do interact with the heated projectile/target and the produced fireball. As a result, a larger number of nucleons participate in the reaction in MUFFIN-SMASH as compared to MCG, at the same value of the impact parameter. This discrepancy becomes smaller with increasing collision energy as the interpenetration process becomes faster, and it becomes negligible at the top RHIC energy.

\subsection{Rapidity distributions}

\begin{figure}
    \centering
    \includegraphics[width=0.5\textwidth]{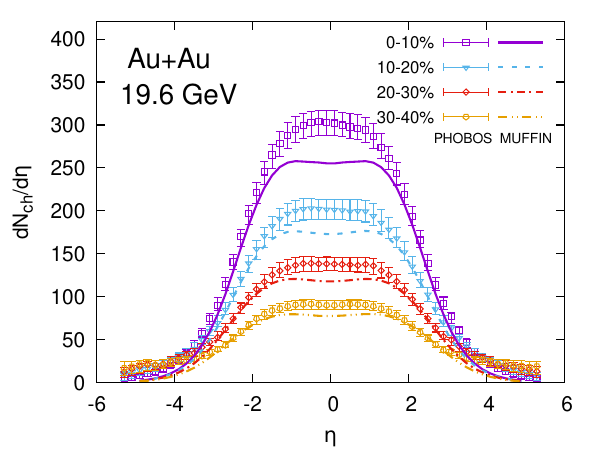}
    \caption{Pseudorapidity distributions of charged hadrons at $\snn=19.6$~GeV Au+Au collisions for various centralities obtained from hybrid three-fluid  model MUFFIN-SMASH compared to the experimental data from PHOBOS collaboration \cite{PHOBOS:2010eyu}.}
    \label{fig:3FH_dndeta_19.6}
\end{figure}

\begin{figure}
    \centering
    \includegraphics[width=0.5\textwidth]{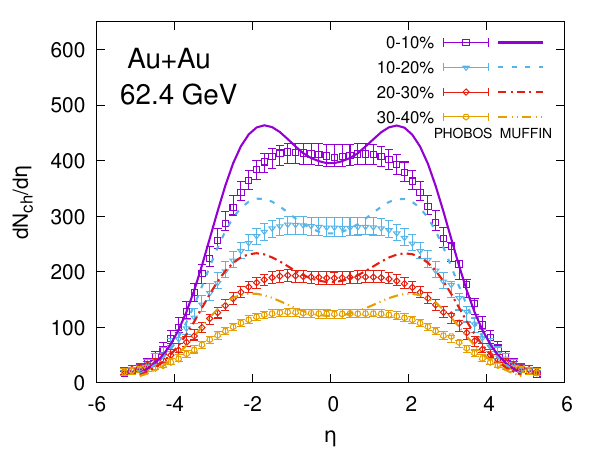}
    \caption{Same as Fig. \ref{fig:3FH_dndeta_19.6}, but for $\snn=62.4$~GeV Au+Au collisions. The experimental data are from PHOBOS collaboration \cite{PHOBOS:2010eyu}.}
    \label{fig:3FH_dndeta_62.4}
\end{figure}

Figures \ref{fig:3FH_dndeta_19.6} and \ref{fig:3FH_dndeta_62.4} show the pseudorapidity distributions of charged hadrons obtained from the hybrid three-fluid  model MUFFIN-SMASH. At $\snn=19.6$~GeV, our model underestimates the multiplicity, mainly in the most central collisions. At $\snn=62.4$~GeV, our model shows a two-peak structure, which is not seen in the experimental data. However, the midrapidity values of multiplicity are reproduced at this energy.
At this point, we note that the $dN_{\rm ch}/d\eta$ values at mid-rapidity from the 3-fluid simulations  are well fitted with the 2-component MCG model, however our fitted values of $n_{\rm pp}$ are slightly different from those used in the 2-component MCG fit to the experimentally measured $dN_{\rm ch}/d\eta$. For example, the values of $n_{\rm pp}$ in our fit are 0.89 and 0.99 for $\snn=7.7$ and $11.5$ GeV, respectively, whereas STAR reports $n_{\rm pp}=1.12$ for $\snn=9.2$~GeV.

\begin{figure}
    \centering
    \includegraphics[width=0.5\textwidth]{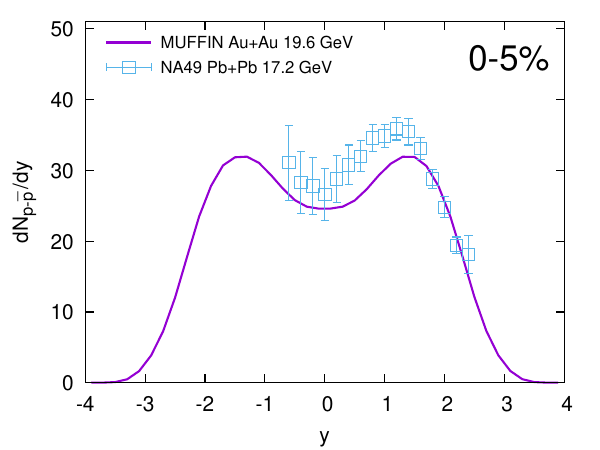}
    \caption{Rapidity distribution of net-protons in 0-5\% Au+Au collisions at $\snn=19.6$~GeV obtained from hybrid three-fluid  model MUFFIN-SMASH compared to the experimental data of Pb+Pb collisions at $\snn=17.2$~GeV from NA49 collaboration \cite{NA49:1998gaz}.}
    \label{fig:3FH_dndy_netproton_19.6}
\end{figure}

\begin{figure}
    \centering
    \includegraphics[width=0.5\textwidth]{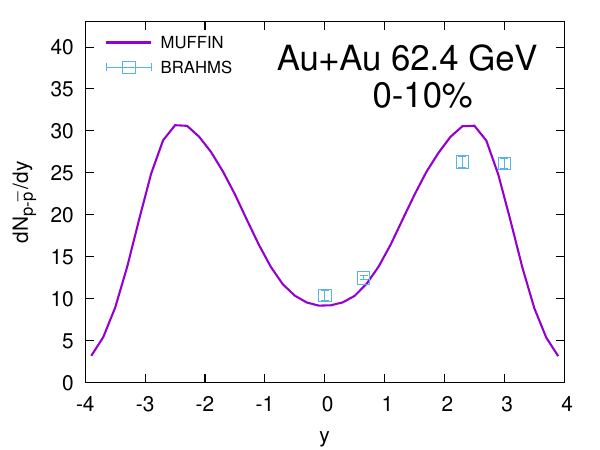}
    \caption{Same as Fig. \ref{fig:3FH_dndy_netproton_19.6}, but for 0-10\% Au+Au collisions at $\snn=62.4$~GeV. The experimental data are from BRAHMS collaboration \cite{BRAHMS:2009wlg}.}
    \label{fig:3FH_dndy_netproton_62.4}
\end{figure}

Figures \ref{fig:3FH_dndy_netproton_19.6} and \ref{fig:3FH_dndy_netproton_62.4} show the rapidity distributions of net-protons obtained from our model. Since there are no experimental data at $\snn=19.6$~GeV, we compare our results with the experimental data for Pb+Pb collisions at $\snn=17.2$~GeV from the NA49 experiment. The slight difference between MUFFIN-SMASH and the measured data is partly caused by different nucleon numbers of collided nuclei. Although there are only four experimental points at $\snn=62.4$~GeV, MUFFIN-SMASH reproduces the shape of the distribution quite well. The consistency of the net-proton rapidity distribution between our model and the experimental data indicates that MUFFIN yields correct baryon stopping. 

The pseudorapidity distributions of charged hadrons at both energies indicate that there could be slightly stronger friction in the model, which would bring more energy to midrapidity. However, that would also result in stronger transverse expansion and stronger baryon stopping, bringing the two peaks in net-proton rapidity distributions closer together, which would worsen the reproduction of the experimentally measured net-proton rapidity distribution.

\subsection{Transverse momentum spectra}

Next, we compute the transverse momentum spectra of $\pi^+$, K$^+$, protons, and antiprotons. The spectra are calculated for $|\rap|<0.1$, weak decays are included in proton and antiproton spectra, and excluded for pion spectra. In order to make the plots more legible, the spectra for different centralities are scaled by different factors.

\begin{figure}
    \centering
    \includegraphics[width=0.5\textwidth]{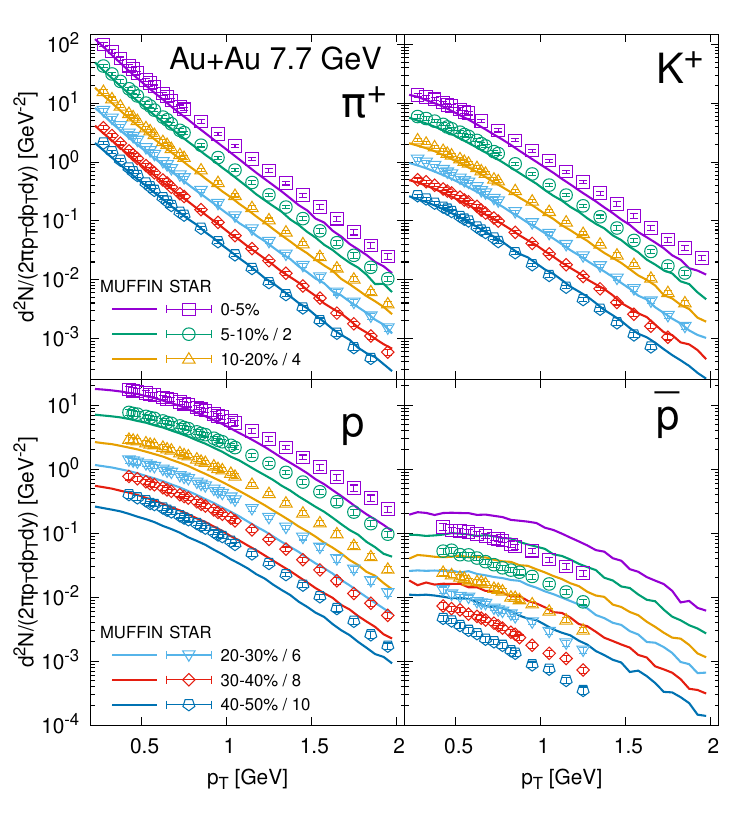}
    \caption{Transverse momentum spectra of positively charged pions (upper left), kaons (upper right), protons (lower left) and antiprotons (lower right) in Au+Au collisions at $\snn=7.7$~GeV for various centralities obtained from the hybrid three-fluid model MUFFIN-SMASH compared to the experimental data from STAR collaboration \cite{STAR:2017sal}.}
    \label{fig:3FH_spectrum_7.7}
\end{figure}

\begin{figure}
    \centering
    \includegraphics[width=0.5\textwidth]{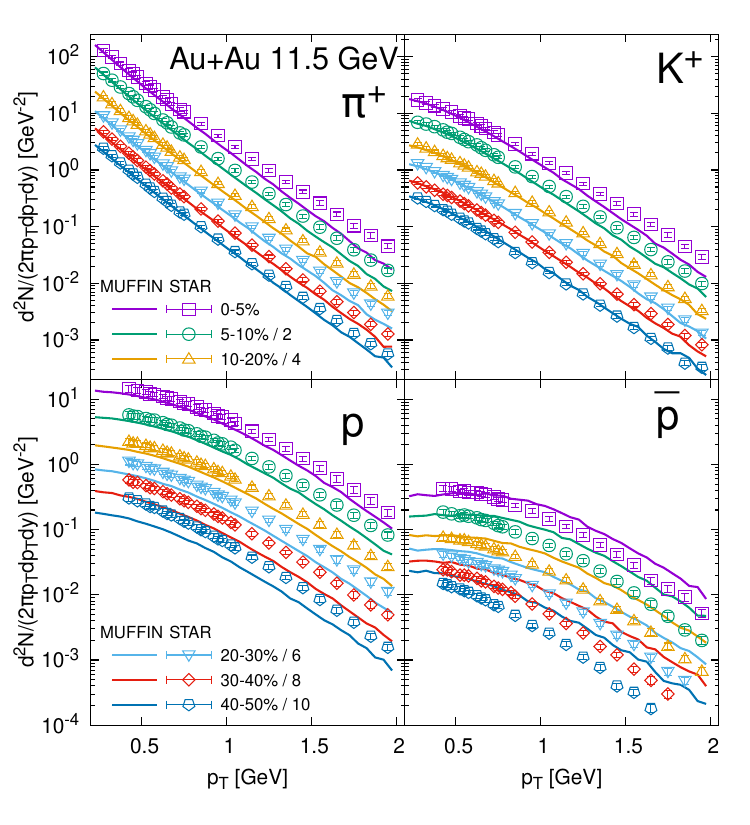}
    \caption{Same as Fig. \ref{fig:3FH_spectrum_7.7}, but for $\snn=11.5$~GeV Au+Au collisions. The experimental data points are from STAR collaboration \cite{STAR:2017sal}.}
    \label{fig:3FH_spectrum_11.5}
\end{figure}

\begin{figure}
    \centering
    \includegraphics[width=0.5\textwidth]{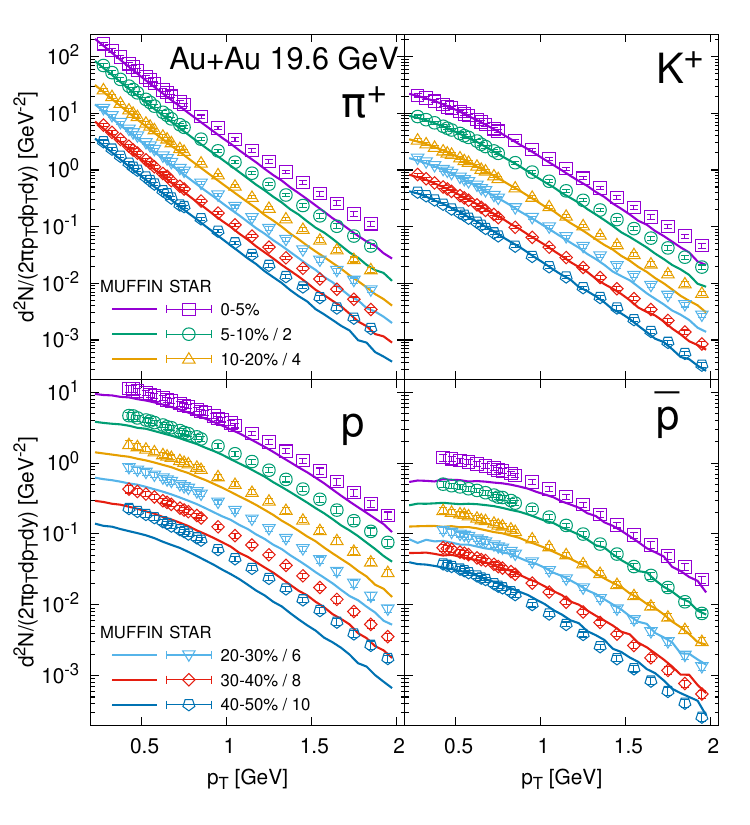}
    \caption{Same as Fig. \ref{fig:3FH_spectrum_7.7}, but for $\snn=19.6$~GeV Au+Au collisions. The experimental data points are from STAR collaboration \cite{STAR:2017sal}.}
    \label{fig:3FH_spectrum_19.6}
\end{figure}

\begin{figure}[ht]
    \centering
    \includegraphics[width=0.5\textwidth]{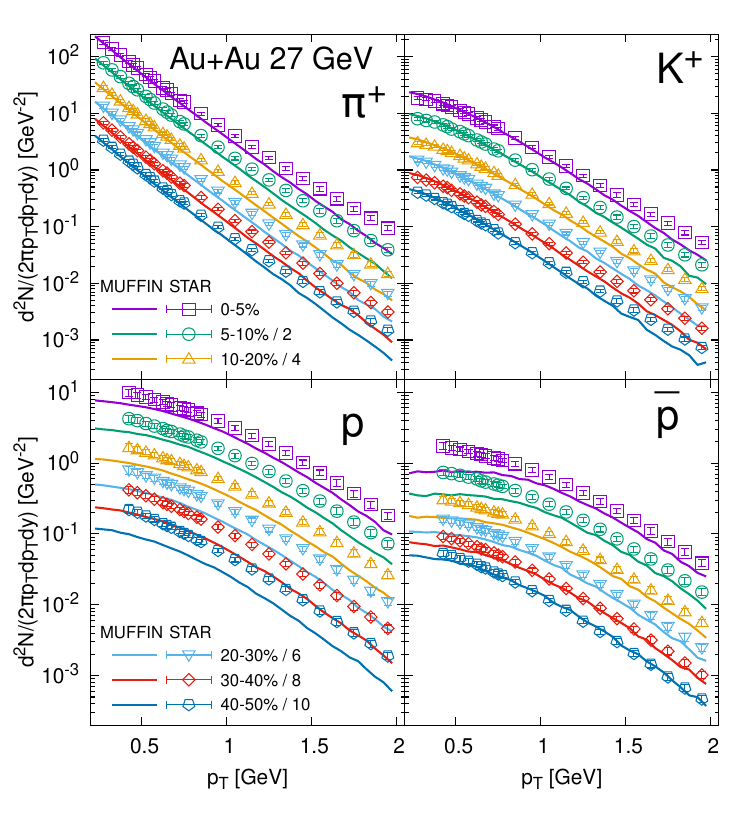}
    \caption{Same as Fig. \ref{fig:3FH_spectrum_7.7}, but for $\snn=27$~GeV Au+Au collisions. The experimental data points are from STAR collaboration \cite{STAR:2017sal}.}
    \label{fig:3FH_spectrum_27}
\end{figure}

\begin{figure}
    \centering
    \includegraphics[width=0.5\textwidth]{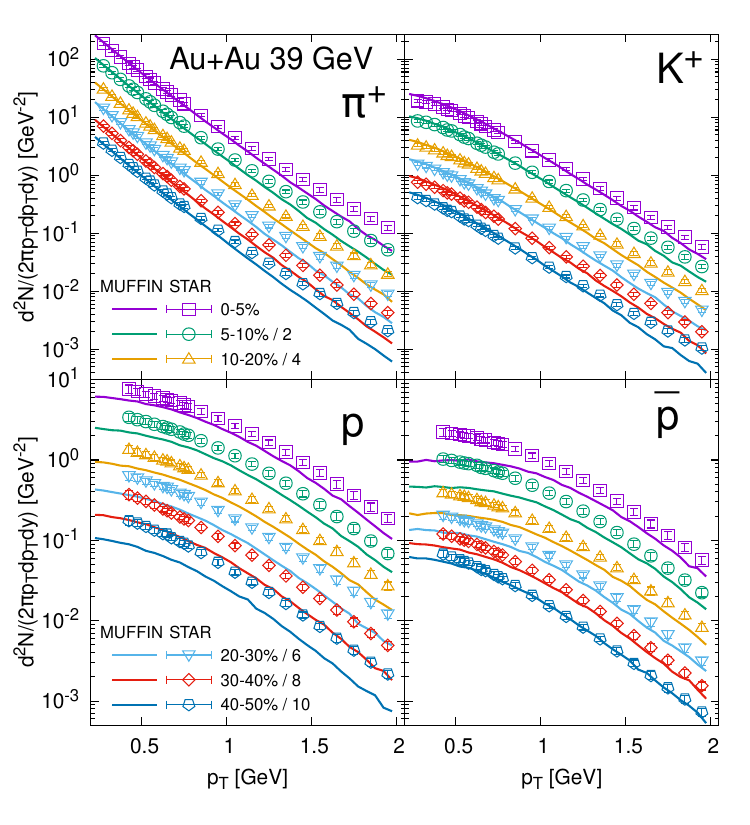}
    \caption{Same as Fig. \ref{fig:3FH_spectrum_7.7}, but for $\snn=39$~GeV Au+Au collisions. The experimental data points are from STAR collaboration \cite{STAR:2017sal}.}
    \label{fig:3FH_spectrum_39}
\end{figure}

\begin{figure}
    \centering
    \includegraphics[width=0.5\textwidth]{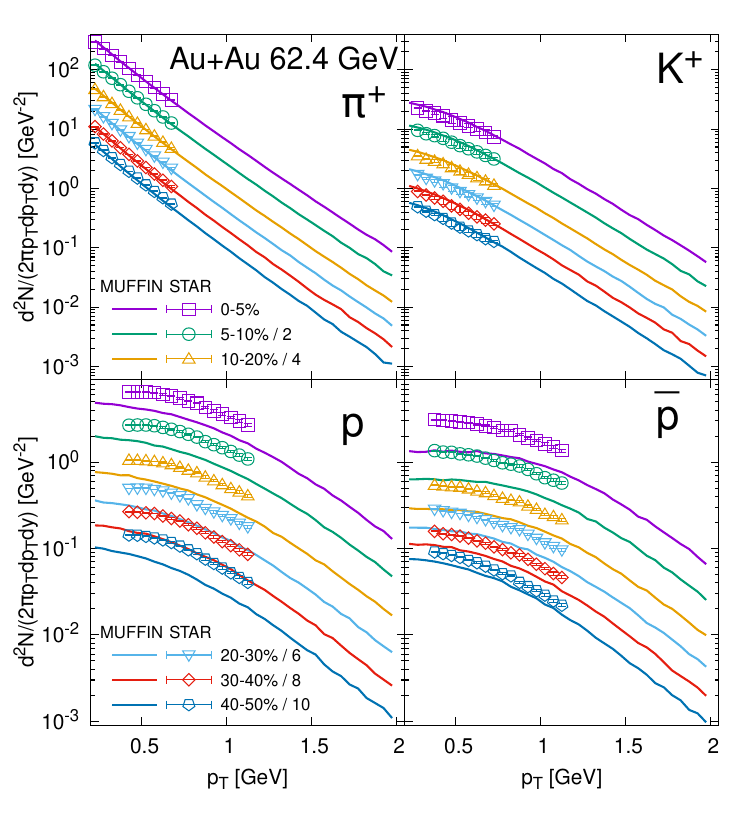}
    \caption{Same as Fig. \ref{fig:3FH_spectrum_7.7}, but for $\snn=62.4$~GeV Au+Au collisions. The experimental data points are from STAR collaboration \cite{STAR:2008med}.}
    \label{fig:3FH_spectrum_62.4}
\end{figure}

At $\snn=7.7$~GeV (Fig. \ref{fig:3FH_spectrum_7.7}), MUFFIN-SMASH reproduces the pion and kaon spetra well, underestimates the proton spectra in particular at mid-central collisions, and overshoots the antiproton spectra.  This indicates some deficit of baryon charge at mid-rapidity, and slightly stronger friction would be needed to fix it; however, at higher energies, this discrepancy disappears. At $\snn=11.5$~GeV (Fig. \ref{fig:3FH_spectrum_11.5}) the results are similar to the lowest energy except for the antiproton spectra, which are closer to the experimental data and even reproduce the low-$p_T$ data for the most central collisions. At $\snn=19.6$~GeV (Fig. \ref{fig:3FH_spectrum_19.6}) the pion spectra start to be underestimated at high-$p_T$. The antiproton spectra at this energy are reproduced for $p_T>1$~GeV. At $\snn=27$~GeV (Fig. \ref{fig:3FH_spectrum_27}), the trend with pion spectra continues. However, antiproton spectra are closer to the data, and for mid-central collisions, they are described perfectly. The results of simulations at $\snn=39$~GeV (Fig. \ref{fig:3FH_spectrum_39}) show the same hierarchy as at $\snn=27$~GeV. At $\snn=62.4$~GeV (Fig. \ref{fig:3FH_spectrum_62.4}), the experimental data are available only  at low-$p_T$. In this range, both pion and kaon spectra agree perfectly with the data, while proton and antiproton spectra are quite underestimated. This, however, cannot be adjusted with the tuning of the friction, because the net-baryon number at midrapidity is correct.

Although not all spectra are reproduced perfectly, the slopes of the spectra in our simulations generally agree with the experimental data, which means that MUFFIN-SMASH generates a correct strength of the collective transverse flow.


\subsection{Anisotropic flow}

\begin{figure}
    \centering
    \includegraphics[width=0.5\textwidth]{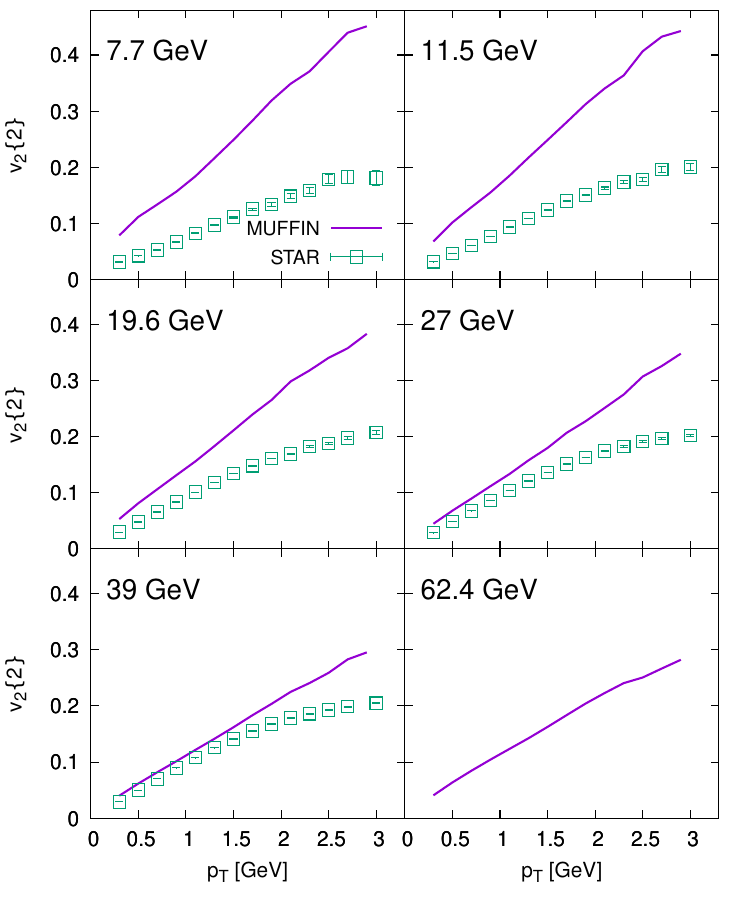}
    \caption{Elliptic flow of charged hadrons as a function of transverse momentum in 20-30\% Au+Au collisions at energies $\snn=7.7-62.4$~GeV obtained from the hybrid three-fluid  model MUFFIN-SMASH compared to the experimental data from STAR collaboration \cite{STAR:2012och}.}
    \label{fig:3FH_v2_pT}
\end{figure}

Finally, we present elliptic flow of charged hadrons, calculated using 2-particle cumulant method \cite{Bilandzic:2010jr}. Figure \ref{fig:3FH_v2_pT} shows the $p_T$-dependent elliptic flow for 20-30\% Au+Au collisions at all studied energies computed from the model and compared to the experimental data from STAR \cite{STAR:2012och}. It is apparent that the $v_2$ obtained from our model is extremely overestimated at low energies. With increasing collision energy, our results are slowly approaching the experimental data, and at $\snn=39$~GeV, there is a near agreement with the experimental $v_2$. Unfortunately, the experimental data at $\snn=62.4$~GeV are not available.  There is clearly room for elliptic flow suppression by shear viscosity, which is not included in this study. The amount of needed suppression grows with decreasing collision energy, which is consistent with an observation made in \cite{Karpenko:2015xea} that the effective ratio of shear viscosity to entropy density of the medium should grow with decreasing collision energy. Here we note that in MUFFIN, certain non-equilibrium effects are taken into account, as the medium when seen as a whole, is not in local equilibrium due to counter-streaming flows of the fluids. However, another kind of non-equilibrium due to finite mean free path, is not present when the perfect-fluid approximation is used.

\begin{figure}
    \centering
    \includegraphics[width=0.5\textwidth]{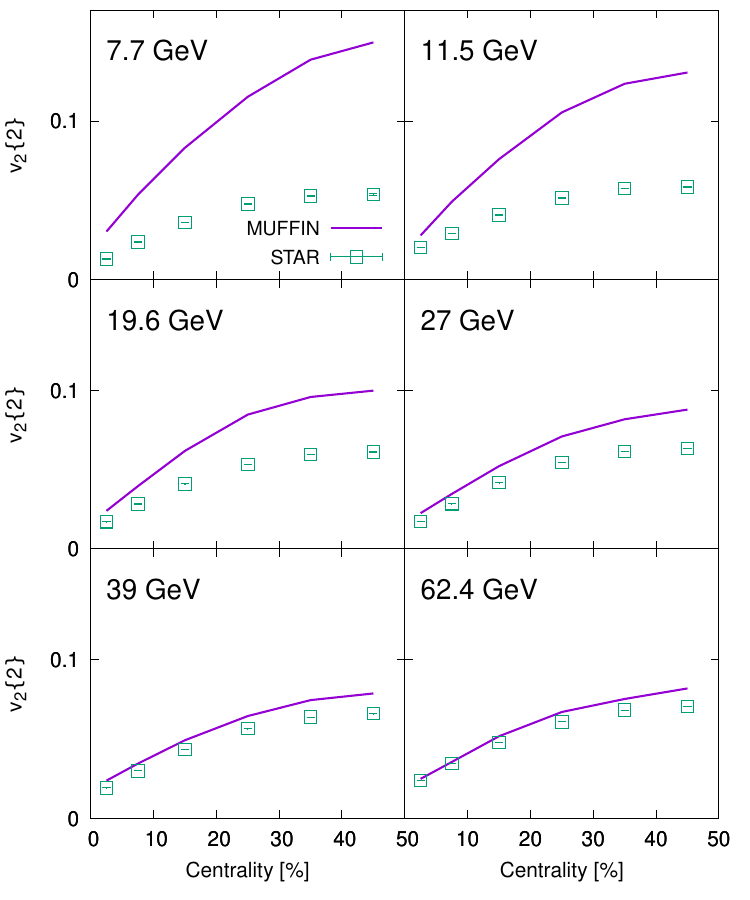}
    \caption{$p_T$-integrated elliptic flow of charged hadrons as a function of centrality in Au+Au collisions at energies $\snn=7.7-62.4$~GeV obtained from the hybrid three-fluid  model MUFFIN-SMASH compared to the experimental data from STAR collaboration \cite{STAR:2012och, STAR:2017idk}.}
    \label{fig:3FH_v2_centrality}
\end{figure}

A similar hierarchy can be seen in the centrality dependence of elliptic flow integrated over $0.2<p_T<2.0$~GeV, shown in Fig. \ref{fig:3FH_v2_centrality}. In this case, the experimental data at the two largest studied energies are only slightly overestimated. This is mainly because the hadron yields decrease with $p_T$, and therefore this observable is not so sensitive to the high-$p_T$ hadrons. We also note that the overestimation of the flow at lower energies grows towards non-central collisions, while the flow in the most central collisions is relatively close to the data.

\begin{table}[tb]
 \centering
 \begin{tabular}{ c | c | c }
 centrality & $b$ [fm] in \cite{Ivanov:2014zqa} & $b$ [fm], this work \\ \hline \hline
 0-5\% & 2.0 & 0 -- 3.2 \\ 
 5-10\% & 4.0 & 3.2 -- 4.57 \\ 
 20-30\% & 6.0 & 6.48 -- 7.95 \\
 30-40\% & 8.0 & 7.95 -- 9.2 \\
 \end{tabular}
\caption{Comparison of impact parameter (ranges) for key centrality classes between \cite{Ivanov:2014zqa} and this work. The rightmost column correspond to impact parameter ranges from Table \ref{tab:centrality_determination} for $\snn=19.6$~GeV.}
\label{tab:impact_parameter_comparison}
\end{table}

The elliptic flow has been previously studied  using three-fluid hydrodynamic model in \cite{Ivanov:2014zqa}. 
Like in this work, the perfect-fluid approximation was employed there, nevertheless the elliptic flow across BES energies was reproduced well at 5-10\% and 20-30\% centralities, and even underestimated in most central collisions.  The most important reason for this discrepancy is in different centrality selection: in \cite{Ivanov:2014zqa}  fixed, integer values of impact parameter were used for each centrality. As shown in Table~\ref{tab:impact_parameter_comparison} the values for mid-central collisions are noticeably lower than the values used in the present study. 
Hence, more central events were simulated effectively in \cite{Ivanov:2014zqa}, which results in a smaller elliptic flow. Moreover, in most central collisions, the main contribution to the elliptic flow is due to fluctuations of the initial state, which is missing in \cite{Ivanov:2014zqa}, and therefore again resulting in weaker elliptic flow.


\section{Conclusions}

We developed a next-generation hybrid three-fluid  model for simulating heavy-ion collisions at energies from few to few tens of GeV. This model is aimed for phenomenological studies of heavy-ion collisions at BES energies at RHIC, NA61/SHINE at CERN and FAIR at GSI. The main features of the model include:
\begin{itemize}\setlength{\itemsep}{-2pt}
    \item fluctuating initial conditions,
    \item hyperbolic coordinate system,
    \item Monte Carlo hadron sampling at particlization
    \item SMASH for hadronic rescatterings,
    \item EoS can be easily changed.
\end{itemize}
The friction terms between the fluids are parametrized in a rather simplistic way following \cite{Ivanov:2005yw}.
As a rigorous derivation of the friction terms from the underlying kinetic theory is lacking, the parametrizations are essentially educated guesses. Therefore, we scaled the friction terms with factors which depend on the center-of-mass energy of the interpenetrating fluid elements, and thus regulated the strength of friction. The scaling factors were then fitted to reproduce available experimental data from RHIC BES for transverse momentum spectra and rapidity distributions. We showed the first results calculated using this model, including rapidity distributions, transverse momentum spectra, and elliptic flow. The model lacks viscous corrections, which results in an overestimation of the elliptic flow. Adding viscosity to the model is among our plans for future studies.

\begin{acknowledgments}

JC, IK and BT acknowledge support by the project Centre of Advanced Applied Sciences, No.~CZ.02.1.01/0.0/0.0/16-019/0000778, co-financed by the European Union, and by the grant GA22-25026S of the Czech Science Foundation. IK acknowledges support by the Ministry of Education, Youth and Sports of the Czech Republic under grant ``International Mobility of Researchers – MSCA IF IV at CTU in Prague'' No.\ CZ.02.2.69/0.0/0.0/20\_079/0017983. BT acknowledges support from VEGA 1/0521/22. Computational resources were supplied by the project ``e-Infrastruktura CZ'' (e-INFRA LM2018140) provided within the program Projects of Large Research, Development and Innovations Infrastructures. PH was supported by the program Excellence Initiative--Research University of the University of Wroc\l{}aw of the Ministry of Education and Science.
\end{acknowledgments}

\bibliographystyle{h-physrev}
\bibliography{main}

\end{document}